\documentclass[10pt,a4paper]{article}
\usepackage{vmargin,color}
\setmarginsrb{2.0cm}{3.0cm}{2.0cm}{1.0cm}{0.0cm}{0.0cm}{0.0cm}{1.0cm}
\usepackage[french,english]{babel}

\usepackage{graphicx,float,setspace,amssymb,amsmath}

\newcommand{\equa}[1]{\begin{eqnarray} \label{#1}} 
\newcommand{\auqe}{\end{eqnarray}} 
\newcommand{\tab}[1]{\begin{tabular}{#1}} 
\newcommand{\bat}{\end{tabular} \\ }

\begin{document}
%
\title {Magnetization in uniaxial spherical nanoparticles: consequence
on the interparticle interaction
}

\author {V. Russier \\ \noindent
ICMPE, UMR 7182 CNRS and Universit\'e Paris Est, \\ \noindent
2-8 rue Henri Dunant 94320 Thiais, France. \\ \noindent
russier@icmpe.cnrs.fr
}
\date{}
\maketitle


\begin{abstract}
We investigate the interaction between spherical magnetic nanoparticles 
which present either a single domain or a vortex structure. 
First the magnetic structure of a uniaxial soft sphere is revisited, and then
the interaction energy is calculated from a micromagnetic simulation.
In the vortex regime the orientation of the vortex relative to the easy axis
depends on both the particle size and the anisotropy constant.
We show that the leading term of the interaction is the dipolar interaction 
energy between the magnetic moments. For particles presenting a vortex 
structure, we show that the polarization due to the dipolar field must be 
included. The parameters entering in the dipolar interaction are deduced from 
the magnetic behavior of the isolated particle. 
\end{abstract}


\section{ Introduction }
\label{intro} 
The magnetic behavior of nanometric particles either isolated or
in nanostructured bulk materials is now quite well undertood both from 
experiments or numerical calculations \cite{rev_skomski,rev_bader}.
With the growing diversity of systems
made of such nano-objects as building blocks either as 2D or 3D systems in non
magnetic environment or in colloidal suspensions as ferrofluids, 
a precise knowledge of interparticles interaction is needed.
This has already been done at least partially in a variety of systems, such as 
nanograins \cite{inter_grain_1,inter_grain_2} nanorings \cite{inter_nanoring} 
or flat nanodots \cite{inter_dot_1,layer_nanodot}.
with a predominant attention paid on short range effects.
A lot of work remains to be done in this field and especially for 
spherical particles. In particular it seems important to developp models
for the long ranged and anisotropic dipolar interaction.
A further interpretation of experimental results as those of \cite{sph_nano1}
necessitates such models.

\section{ Magnetization structure and hysteresis } 
\label{part_1} 
\begin{figure}[t]
\includegraphics[width = 8.0 cm]
{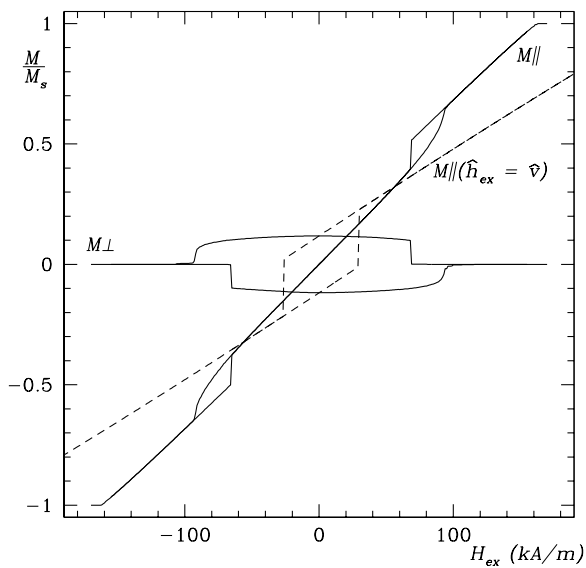}
\caption{ \label{hyst_r45}
Magnetization parallel and normal to the field.
$\hat{h}_{ex}$ = $\hat{a}$ (solid line) or $\hat{v}$ (dashed line).
$R = 45nm$, $K_1 = 3.10^4 J/m$. 
}
\end{figure}

We consider a spherical particle characterized by a radius $R$ ranging from 
$10 nm$ to $45 nm$, an  exchange constant, $A_{x} = 1.10^{-11}J/m$ and a 
saturation magnetization $J_s = 1T$ corresponding roughly to Permalloy. 
The anisotropy constant of the uniaxial magneto-crystalline energy is varied 
between $K_1 = 0$ and $7.10^4 J/m^3$. 
The numerical calculations are performed with the
the framework of micromagnetism with the code MAGPAR \cite{magpar} 
based on a FEM method.
The particle volume is $v_s$ and hatted letters denote unit
vectors. Small particles, up to roughly $R = 20 nm$ for our parameters, are 
uniformly magnetized as single magnetic domains with square hysteresis curve.
When the particle radius increases beyond roughly $R = 20 nm$,
a vortex structure is obtained, and the local magnetization profile,
$\vec{m}({\vec r})$, is decomposed in its cylindrical components using the 
vortex axis, say $\hat{v}$, as the cylindrical axis. 
At zero or small values of the external 
field, the direction of the vortex relative to the easy axis, $\hat{a}$, 
depends on both the value of $R$ and of $K_1$, a behavior already obtained 
in the case of the cubic anisotropy \cite{calc_sph_c}.
We find, as expected, 
$\hat{v}$ $\perp$ $\hat{a}$ either for $R > R_{th}$ or $K_1 > K_{1, th}$ 
at constant $K_1$ or $R$ respectively, $R_{th}$ and $K_{1, th}$ being some 
threshold values. 
($R_{th} = 26 nm$ for $K_1 = 3.10^4 J/m^3$ and $K_{1, th} = 2.10^3 J/m^3$ for
$R = 45nm$).
The orientation of $\hat{v}$ relative to $\hat{a}$ affects
strongly the $M(H_{ex})$ curve. Whith an external field along
$\hat{a}$, $\hat{v}$ $\perp$ $\hat{a}$ leads to a magnetization curve 
qualitatively similar to that of a flat nanodot with an in plane external
field \cite{nanodot,rev_bader}: 
threre is no remanence and $M(H_{ex})$ presents two lobes. 
For spherical particles these lobes are associtated to the rotation of
the vortex core in the direction of the field prior to its annihilation.
The magnetization normal to the field takes a non zero constant 
value corresponding to the vortex core magnetization, when the 
variation of $M\parallel$ results from a shift of the vortex normal to the 
field. this value coincides with the remanence obtained with
$\hat{h}_{ex}$ = $\hat{v}$.  This behavior is displayed on figure \ref{hyst_r45}.

As is generally obtained in nanodots or spherical particles 
\cite{nanodot,calc_sph_1,calc_sph_2} the 
magnetization $M$ in the direction of the external field depends
nearly linearly on the field in the vicinity of $H_{ex}$ = 0 and 
away from swhitching points where the vortex reverses as a whole. 
Such a linear behavior is observed both when
$\hat{h}_{ex}$ = $\hat{v}$ or $\hat{h}_{ex}$ $\perp$ $\hat{v}$. 
This means that the susceptibility $\chi$ defined as 
$\frac{\partial M}{\partial H_{ex}} = \chi$
does not depend on the field to a very good approximation. 
This can be exploited to obtain the variation
of the total energy with respect to the external field.
We analyse the variation of the magnetization, $\Delta M$ 
as the polarization of the sphere induced by the field. 
Starting from $\Delta M(H_{ex})$ = $\chi H_{ex}$, we deduce 
$\Delta M(H_{ex})$ from the energy, $E(H_{ex})$ by writting an equilibrium 
equation
$\frac{\partial E(\Delta M)}{\partial \Delta M} = 0$.
The total energy depends explicitly on $H_{ex}$ through the Zeeman term, 
$E_Z = -\mu_0 H_{ex}(m(0)\hat{v}.\hat{h}_{ex} + \Delta M)$, 
where we have expressed 
the permanent magnetization in the absence of the field as 
$\vec{M}(H_{ex} = 0)$ 
= $m(0)\hat{v}$ and $m(0)$ denotes the magnitude of 
the vortex core magnetization in the absence of the field. 
Then we get
\equa{pol_ener2}
\frac{\partial}{\partial \Delta M}\left( E_{dm} + E_{x} + E_{a} \right ) =
\mu_0 H_{ex} 
\auqe
Therefore the variation of the total energy is 
\equa{pol_ener3}
E(H_{ex}) - E(0) = \frac{\mu_0}{2 \chi}\Delta M^2 + E_Z(H_{ex}) 
\auqe
where we have used $\Delta M(H_{ex})$ = $\chi H_{ex}$. 
$\mu_0$ is the vacuum permeability and $E_Z$ the Zeeman energy.
Notice that (\ref{pol_ener2}) is exact 
while (\ref{pol_ener3}) holds only in the case of a linear dependence of
$\Delta M(H_{ex})$ with respect to $H_{ex}$.
The first term of the r.h.s. of (\ref{pol_ener3})  
is the polarization energy of the sphere \cite{polar} and 
corresponds to the energy cost for the reorientation of the magnetization 
inside the sphere. It can be written equivalently as $(\mu_0\Delta M H_{ex})/2$ 
%
\section{ Interaction between magnetic nanospheres } 
\label{part_2} 
The interaction energy between two magnetic nanoparticles in terms of the 
interparticle distance, $r_{12}$, is defined from the total energy of the two 
spheres system
\equa{delta_e}
E_{int}(1,2) = E_{tot}(1,2) - E_{tot}(r_{12} \rightarrow \infty) 
\auqe
where $(1,2)$ is a short notation for the orientation and location variables 
of the particles. We expect to get a form dictated by the dipolar interaction 
between the magnetic moments of the approaching spheres wich reads
\equa{inter_dip}
& E_{dip} = 
\frac{\mu_0 m_1 m_2}{4 \pi r_{12}^3} d_{112}(\hat{m}_1, \hat{m}_2, \hat{r}_{12})
& \\ \label{d_112}
& d_{112}(\hat{m}_1, \hat{m}_2, \hat{r}_{12}) = 
\hat{m}_1.\hat{m}_2 - 3(\hat{m}_1.\hat{r}_{12})(\hat{m}_2.\hat{r}_{12})
&
\auqe
where $m_i$ are the magnitude of the magnetic moments.
For single domain particles, $m_i = M_s v_s$ where $M_s$ is the saturation 
magnetization of the particles, and the orientations $\hat{m}_i$ result from 
the minimum of $d_{112}$ given in (\ref{d_112}). For particles without 
magnetocristalline anisotropy, this gives: $\hat{m}_1$ = $\hat{m}_2$
= $\hat{r}_{12}$ and $d_{112}$ = -2. For non zero
magnetocristalline energy on both particles with easy axes 
$\hat{a}_i$, the orientations $\hat{m}_i$ result of the interplay 
between the anisotropy energy tending to align $\hat{m}_i$ on $\hat{a}_i$ 
and the energy (\ref{inter_dip}) tending to minimize the angular function 
(\ref{d_112}). If $K_1$ takes a non vanishing value only 
in one particle say $i = 1$ and is large enough to impose 
$\hat{m}_1$ = $\hat{a}_1$, $\hat{m}_2$ must orient in the dipolar field due 
to particle $1$ {\it i.e.} in such a way that $d_{112}$ =
$\hat{m}_2.\hat{a}_1 - 3(\hat{m}_2.\hat{r}_{12})(\hat{a}_1.\hat{r}_{12})$
is minimum. 
The whole behavior outlined above is very well reproduced by the full 
micromagnetic calculation which demonstrates the dipolar nature of the 
interaction between single domain particles.

In the case of particles large enough to present a vortex 
structure, the orientations of the effective moments of the 
particles are the vortex directions, $\hat{v}_i$, and the values of the
moments correspond to the vortex cores magnetizations. 
We introduce the coefficient $\alpha_i$ = $m_i/(M_s v_s)$.
The value taken by $\alpha$ depends on both the 
characteristics of the particle and $r_{12}$ and $d_{112}$
through the polarization of the particle by the dipolar field of the 
second one. 
($\alpha(1,2)$ = $\alpha(r_{12},d_{112})$ and $\alpha_0 = \alpha(\infty)$). 
A simple approximation for the interaction energy 
is built by considering that
each particle is in the dipolar field of the other. We have to take into
account two contributions. The first one is given by (\ref{inter_dip}),
and the second one is twice the polarization energy 
of one sphere in the field of the second one. The second contribution has been
introduced in (\ref{pol_ener3}) for one particle in a constant external field.
The role of $m(o)$ is played by $M_s v_s \alpha_0$
while the induced moment in the direction of the dipolar field is 
$
\vec{p} = p\hat{h}_{dip} = \chi H_{dip}(r_{12})\hat{h}_{dip}.
$
We consider the case where the vortex $\hat{v}_i$ is free to orient in the 
direction of the dipolar field due to particle {\it j $\ne$ i}. This 
corresponds to either the absence of anisotropy or particles large enough for 
the vortex to be normal to the easy axis and $\hat{a}_1$ = $\hat{a}_2$.
In this case we have
$
\vec{p} = (\alpha(1,2) - \alpha_0))M_s v_s \hat{h}_{dip} = 
\Delta\alpha(1,2) M_s v_s \hat{v} 
$
and adding twice the first term of (\ref{pol_ener3}) to the dipolar
energy we get 
\equa{inter_ener}
 E_{int}(1,2) = 
 \frac{\mu_0 (M_s v_s)^2}{4 \pi r_{12}^3}
 \alpha_0(\alpha_0 + \Delta\alpha(1,2))
d_{112}  
\auqe
which coincides with the interaction energy between polar polarizable hard 
spheres. Then we have to calculate $\Delta\alpha(1,2)$; for the simple case
of two particles, introducing $u$ = $\chi/(4_\pi R^3)$ we get
\equa{delta_alpha}
\Delta\alpha(1,2) = \alpha_0 \frac{-u d_{112}}{(r_{12}/R)^3 + u d_{112}}
\auqe

\begin{figure}[t]
\includegraphics[width = 8.0 cm]
{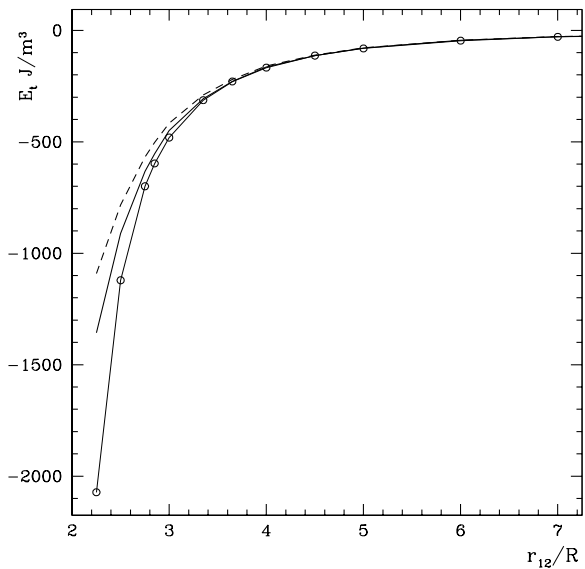}
\caption{ \label{ener_r35_k0}
Interaction energy per unit volume between two spheres. 
$R=35nm$; $K_1 = 0$; $d_{112} = -2$.
Open circles: full calculation, from (\ref{delta_e}) 
(the thin line is a guide to the eye);
solid line: equ.(\ref{inter_ener});
dashed line: simple dipolar approximation, $u = 0$.
}
\end{figure}

A typical example of the interaction in terms of the distance $r_{12}$
for  $d_{112} = -2$ is displayed on figure \ref{ener_r35_k0}. Similar results 
are obtained for other particles. The full calculation is compared to the 
analytical model given by (\ref{inter_ener}). We consider two levels of 
approximation: either the simple dipolar interaction, where the polarization 
energy is neglected ($u = 0$) and the interaction corresponding to the dipolar 
polarizable spheres. Except in the vicinity of contact, ($r_{12} = 2R$)
equ. (\ref{inter_ener}) reproduces quite well $E_{int}(1,2)$. 
The parameters needed, $\alpha_0$ and $\chi$ are deduced from the magnetization 
curve of the isolated particle. We can therefore conclude that the interaction 
between spherical soft magnetic particles is of dipolar nature, and that the 
polarizability must be included when the particles present a vortex structure. 

\section{Concluding remarks}
We can conclude that the leading interaction between soft magnetic
nanoparticles corresponds as expected to the magnetic dipolar interaction. 
Then two situations must be distinguished according to the particle size:
the interaction corresponds to the total magnetic moment without
any polarizability contribution in the case of small particles which
are in a single domain state when isolated, while a polarizability term
must be included when the particles present a vortex structure. In this 
latter case, the value of the permanent moment, represented by the coefficient
$\alpha_0$ in the present work, is no more trivial, but can be determined
as well as the relevant susceptibility $\chi$ from the isolated particle
properties.
We emphasize that here, the solvation od the equation for the variation of
$\Delta\alpha$ in terms of $r_{12}$ and $d_{112}(1,2)$ is quite obvious
since there is only two particles, but this should be not the case for an 
assembly including a large number of particles where the interaction will 
present a $n$-body character. Finaly, because of the the dipolar nature of the
leading interaction, we can confirm the behavior observed experimentaly in
\cite{sph_nano1} since the vortex of neighboring particles are expected to 
allign themselves.
\section*{Acknowledgements}
Fruitful discussions with Dr. F. Mazaleyrat and Dr. Y. Champion are acknowledged.

\end{document}